\documentclass[a4paper,11pt]{article}
\usepackage[top=25mm, bottom=25mm, left=25mm, right=25mm]{geometry}
\usepackage{authblk}

\usepackage{cite}
\usepackage{amsmath,amssymb,amsfonts}
\usepackage{algorithmic}
\usepackage[dvipdfmx]{graphicx}
\usepackage{textcomp}
\usepackage{makecell}
\usepackage{xcolor}

\usepackage{cite}
\usepackage{amsmath,amssymb,amsfonts}
\usepackage{algorithmic}
\usepackage[dvipdfmx]{graphicx}
\usepackage{textcomp}
\usepackage{makecell}

\usepackage{xcolor}
\usepackage{makecell}
\def\BibTeX{{\rm B\kern-.05em{\sc i\kern-.025em b}\kern-.08em
    T\kern-.1667em\lower.7ex\hbox{E}\kern-.125emX}}
\begin{document}

\title{Misophonia Trigger Sound Detection on Synthetic Soundscapes Using a Hybrid Model with a Frozen Pre-Trained CNN and a Time-Series Module\\
}
\author[1]{Kurumi Sashida}
\author[1,2]{Gouhei Tanaka}

\affil[1]{Department of Computer Science, Nagoya Institute of Technology, Nagoya 466-8555, Japan}
\affil[2]{International Research Center for Neurointelligence, The University of Tokyo, Tokyo 113-0033, Japan}
\affil[ ]{\textit{k.sashida.344@stn.nitech.ac.jp, gtanaka@nitech.ac.jp}}

\maketitle

\begin{abstract}
Misophonia is a disorder characterized by a decreased tolerance to specific everyday sounds (trigger sounds) that can evoke intense negative emotional responses such as anger, panic, or anxiety. 
These reactions can substantially impair daily functioning and quality of life. 
Assistive technologies that selectively detect trigger sounds could help reduce distress and improve well-being.
In this study, we investigate sound event detection (SED) to localize intervals of trigger sounds in continuous environmental audio as a foundational step toward such assistive support.
Motivated by the scarcity of real-world misophonia data, we generate synthetic soundscapes tailored to misophonia trigger sound detection using audio synthesis techniques. Then, we perform trigger sound detection tasks using hybrid CNN-based models.
The models combine feature extraction using a frozen pre-trained CNN backbone with a trainable time-series module such as gated recurrent units (GRUs), long short-term memories (LSTMs), echo state networks (ESNs), and their bidirectional variants.
The detection performance is evaluated using common SED metrics, including Polyphonic Sound Detection Score 1 (PSDS1).
On the multi-class trigger SED task, bidirectional temporal modeling consistently improves detection performance, with Bidirectional GRU (BiGRU) achieving the best overall accuracy. Notably, the Bidirectional ESN (BiESN) attains competitive performance while requiring orders of magnitude fewer trainable parameters by optimizing only the readout. We further simulate user personalization via a few-shot ``eating sound'' detection task with at most five support clips, in which BiGRU and BiESN are compared. In this strict adaptation setting, BiESN shows robust and stable performance, suggesting that lightweight temporal modules are promising for personalized misophonia trigger SED.
\vspace{1em} 
\par
\end{abstract}

\section{Introduction}
Misophonia is a disorder of decreased tolerance to specific sounds or associated cues (triggers)\cite{swedo2022consensus}. Triggers are often repetitive and most commonly orofacial sounds (e.g., eating, sniffing, or breathing), but can also include sounds such as keyboard typing \cite{swedo2022consensus,ozuer2025understanding}.
Exposure to trigger sounds can evoke intense negative emotions (e.g., anger, anxiety, or panic) accompanied by autonomic arousal, sometimes described as an involuntary fight-or-flight response\cite{edelstein2013misophonia}.
Misophonic reactions are often context-dependent and modulated by the source identity. For instance, individuals typically react to sounds produced by others, whereas the same sounds produced by themselves are tolerated.
They are not determined by loudness alone \cite{swedo2022consensus}.
Misophonia can lead to restricted social participation and reduced quality of life, stemming from impairments in academic and occupational functioning and the deterioration of interpersonal relationships \cite{swedo2022consensus}.
To cope with these triggers in daily life, individuals employ measures such as leaving the situation, reducing auditory input with earplugs or noise-cancelling headphones, or masking them with alternative sounds such as music or white noise \cite{ozuer2025understanding}.
While these approaches may provide immediate relief, persistent avoidance may increase arousal and awareness of triggers, contributing to symptom maintenance \cite{dozier2023novel}.
Moreover, evidence from research on related sound intolerance conditions suggests that sustained reduction in auditory input (e.g., overuse of hearing protection) can alter loudness perception, consistent with adaptive changes in central auditory gain \cite{hutchison2023acoustic,cherri2024counseling}.
Habituation to trigger sounds should not be assumed in misophonia, as symptoms may remain persistent and, in some cases, may worsen over time\cite{dozier2023novel}. 
Therefore, there is a need for assistive technologies capable of selectively detecting and, as necessary, attenuating or suppressing trigger sounds while maintaining access to the surrounding acoustic environment.
To achieve such selective real-time suppression, a detection mechanism capable of identifying trigger onset, offset, and class is required.

\begin{figure}[!t]
  \centering
  \includegraphics[width=1.0\linewidth]{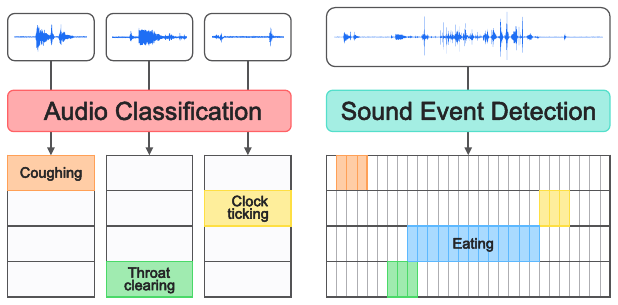}
  \caption{Overview of audio classification and sound event detection (SED). Left: audio classification assigns clip-level labels without temporal localization. Right: SED aims to identify event categories and estimate their onset and offset times within an audio clip.}
  \label{fig:method}
\end{figure}

Prior work on misophonia has mainly characterized responses to pre-selected auditory stimuli and curated stimulus sets \cite{hansen2021sound,orloff2023curation}.
More recently, learning-based approaches have also been explored for recognizing or classifying misophonia-related sounds \cite{bahmei2023misophonia,marosvolgyi2025passt,wunrow2024selective}.
However, audio classification is typically a clip-level task that predicts a class (or multiple classes) for a sound clip ranging from several to over ten seconds; it does not directly identify the specific occurrence intervals (i.e., onset and offset times) within continuous audio (Fig. \ref{fig:method}).
Detecting when and what trigger sounds occur in continuous, everyday audio remains relatively underexplored.

We therefore investigate Sound Event Detection (SED) to estimate both the trigger sound class and its temporal occurrence as a basis for misophonia-oriented assistive applications.
In particular, to realize personalized and real-time misophonia alleviation technologies in the future, it is essential to develop efficient SED models capable of on-device operation with limited computational resources.
SED aims to identify what sound events occur and to localize their onset and offset times in an audio recording \cite{mesaros2021sound}.
In SED research, centering on the Detection and Classification of Acoustic Scenes and Events (DCASE) Challenge Task 4, benchmarks often complement real recordings with synthetic soundscapes containing strong annotations at scale to address the scarcity of real-world data with strong labels \cite{turpault2019sound}.
Specifically, the Domestic Environment Sound Event Detection (DESED) dataset includes real domestic recordings sourced from AudioSet \cite{gemmeke2017audio} as well as synthetic clips generated with Scaper \cite{salamon2017scaper}, for which precise event timestamps are available for supervised training \cite{turpault2019sound}.
As a model framework, the convolutional recurrent neural network (CRNN) has often been adopted in DCASE challenges \cite{turpault2019sound}. The CRNN combines convolutional neural networks (CNNs) for time–frequency pattern extraction with recurrent neural networks (RNNs) and their variants for temporal context modeling.

Notably, misophonia trigger sounds are diverse and vary across individuals, making it challenging to collect strongly labeled real-world recordings that adequately cover the target trigger categories \cite{swedo2022consensus,vitoratou2021item}.

In this study, following the synthesis procedure of DESED/DCASE Task 4\cite{turpault2019sound,dcase20_task4_code}, we generate strongly labeled misophonia trigger soundscapes using Scaper\cite{salamon2017scaper} and train a multi-class SED model in a supervised manner. To ensure a lightweight model architecture, we freeze the pre-trained CNN backbone and train the sequential module. We evaluate the discriminative performance and the temporal localization of event boundaries for closely related orofacial categories (e.g., chewing, sniffing, and breathing) as well as short, impulsive events such as coughing, throat clearing, typing, and clock ticking.

We perform numerical experiments on misophonia trigger SED tasks using a hybrid model that combines a frozen pre-trained CNN feature extractor and a trainable time-series module. As temporal modules, we adopt gated recurrent units (GRUs), long short-term memories (LSTMs), echo state networks (ESNs), and their bidirectional variants.
First, we perform a multi-class trigger SED task on synthetic soundscapes containing seven trigger categories. Our results demonstrate that bidirectional modeling significantly improves performance, with the Bidirectional GRU (BiGRU) achieving the highest accuracy, while the Bidirectional ESN (BiESN) offers a compelling trade-off between performance and parameter efficiency.
Second, addressing the need for individual variability in triggers, we simulate a few-shot user-adaptive Trigger SED task focusing on the ``eating sound'' class. In this data-scarce regime with at most five support clips, the BiESN demonstrates superior stability compared to a fully trainable BiGRU, suggesting its potential for robust personalization in limited-data scenarios.


\section{Dataset Synthesis}
\label{sec:dataset}
Standardized and publicly available datasets for misophonia trigger sound detection remain scarce.
We construct a strongly labeled synthetic soundscape dataset for supervised multi-class SED with seven misophonia-related trigger categories: five human-produced oronasal/throat sounds (eating, sniffing, throat clearing, coughing, and breathing sounds) and two repetitive non-oronasal sounds (typing as a human-generated mechanical sound and clock ticking as a machine-generated mechanical sound), selected based on prior reports of common triggers and practical data availability \cite{vitoratou2021item,hansen2021sound}.
Following the DESED/DCASE Task 4 recipe, we synthesize 10-s soundscapes using the Scaper soundscape synthesis library by mixing foreground events onto background recordings. Strong labels (onset/offset) are obtained directly from the synthesis metadata \cite{turpault2019sound, salamon2017scaper}.

Foreground event candidates are collected from the public corpora including the Free Open-Access Misophonia Stimuli (FOAMS) database \cite{orloff2023curation}, the Misophonia Audiovisual Trigger Archive (MATA) \cite{oh2025chewing}, the Freesound Dataset 50K (FSD50K) \cite{fonseca2021fsd50k}, the Environmental Sound Classification dataset (ESC-50) \cite{piczak2015esc}, and VocalSound \cite{gong2022vocalsound}, while background recordings are drawn from the Sound in the Noise database \cite{dekkers2017sins}, the Tampere University of Technology Acoustic Scenes 2017 dataset \cite{mesaros2016tut}, and the Music, Speech, and Noise corpus \cite{snyder2015musan}, after excluding clips likely to contain any target classes.

Because the source clips may include other sounds (e.g., speech or handling noise), we curate foreground samples in two stages: (i) we apply Yet another Audio MobileNet Network (YAMNet) \cite{yamnet_tfhub_tutorial} to screen candidate clips for each target class, and (ii) we manually audit the clips that YAMNet failed to classify as the target, retaining them unless they are clearly audible as a distinct, non-target event.
The number of curated source clips per class and per dataset is listed in Table~\ref{tab:synth_stats}. During synthesis, we randomly excerpt event segments from longer recordings and apply pitch shifting uniformly in the range of $-3$ to $+3$ semitones.
For the clock ticking class only, we apply spectral-gating noise reduction (noisereduce) to mitigate the stationary hiss between ticks.
Before synthesis, we split the curated source clips into train/validation/test splits with a 3:1:1 ratio, ensuring that no source recording appears in more than one split.
We then synthesize 10-s soundscapes from each split, yielding 6000/2000/2000 clips for train/validation/test, respectively.

\begin{table}[t]
\caption{Statistics of the synthesized misophonia dataset. Source Files indicates the number of unique isolated recordings used for synthesis, Synthesized Clips denotes the number of generated soundscapes containing the class, and Total Duration represents the cumulative length of all event instances for each class.}
\label{tab:synth_stats}
\centering
\renewcommand{\arraystretch}{1.1}
\setlength{\tabcolsep}{4pt}
\begin{tabular}{lrrrr}
\hline
\textbf{Class} & \textbf{Source} & \textbf{Synthesized} & \textbf{Total} & \textbf{Total Duration} \\
& \textbf{Files} & \textbf{Clips} & \textbf{Events} & \textbf{(min)} \\
\hline
Breathing        & 353 & 2962 & 3706 & 106.9\\
Coughing         & 6020 & 3536 & 5111 & 66.1\\
Eating           & 842 & 2468 & 3431 & 76.1\\
Sniffing         & 4111 & 3674 & 5667 & 60.5\\
Throat clearing  & 3480 & 3376 & 4897 & 62.8\\
Typing           & 630 & 2141 & 2746 & 87.2\\
Clock ticking    & 168 & 1848 & 2008 & 138.8\\
\hline
\textbf{Total}   & 15604 & 10000 & 27566 & 598.3\\
\hline
\end{tabular}
\end{table}

\section{Methods}
\label{sec:methods}
\subsection{Overview: A Hybrid Model}
\begin{figure*}[!t]
  \centering
  \includegraphics[width=\linewidth]{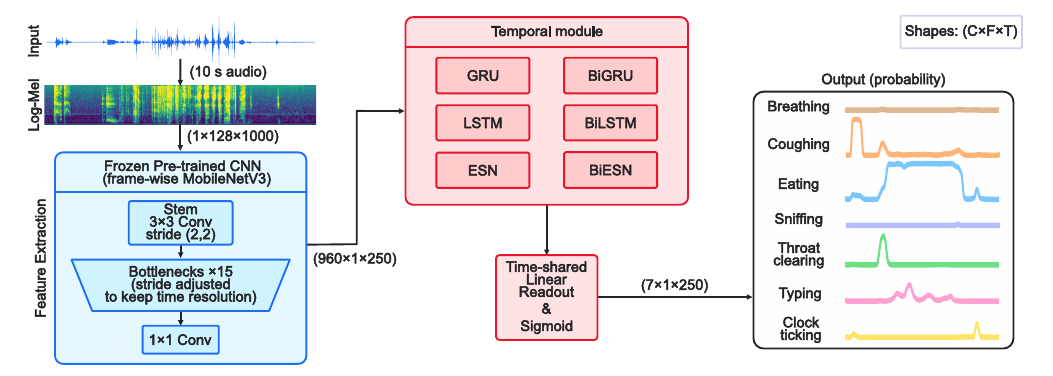}
  \caption{Frozen CNN + temporal module pipeline for frame-wise multi-label sound event detection.
A pre-trained frame-wise MobileNetV3 backbone extracts embeddings $\mathbf{z}_t$, which are processed by a temporal module (GRU/LSTM/ESN; uni- or bidirectional variants) and a time-shared linear + sigmoid readout to yield per-frame class posteriors.}
  \label{fig:model}
\end{figure*}

Fig.~\ref{fig:model} summarizes the end-to-end pipeline of the hybrid models employed in this study. We freeze the pre-trained CNN backbone and train only the temporal module and the shared linear readout to isolate the effect of temporal modeling under a fixed front-end.
To systematically evaluate the impact of temporal modeling on polyphonic SED, we adopt a unified perspective on RNN variants.
While CRNNs typically treat the convolutional front-end and recurrent back-end as a monolithic trainable entity \cite{mesaros2021sound}, our goal is to isolate the contribution of temporal dynamics under resource-constrained settings.

We formulate the temporal modules (GRU, LSTM, ESN) under a unified framework of Generalized Gated State Dynamics, while employing a Linear layer as a non-temporal baseline.
Let $\mathbf{z}_t$ be the input feature at time $t$ extracted by the CNN, and $\mathbf{s}_t$ be the internal state of the temporal module.
We define the general update rule as a balance between retaining past context and integrating new information as follows:
\begin{align}
    \text{Update:} \quad \mathbf{s}_t &= \mathbf{\Gamma}_r \odot \mathbf{s}_{t-1} + \mathbf{\Gamma}_u \odot \tilde{\mathbf{s}}(\mathbf{z}_t, \mathbf{h}_{t-1}), \label{eq:unified_update} \\
    \text{Output:} \quad \mathbf{h}_t &= \mathbf{\Gamma}_o \odot \phi(\mathbf{s}_t), \label{eq:unified_output}
\end{align}
where $\mathbf{h}_t$ is the exposed hidden state, $\tilde{\mathbf{s}}$ is the candidate state derived from the input and previous hidden state, and $\phi$ is an activation function.
The vectors $\mathbf{\Gamma}_r$, $\mathbf{\Gamma}_u$, and $\mathbf{\Gamma}_o$ represent the \textit{retention}, \textit{update}, and \textit{output} factors, respectively, and $\odot$ denotes element-wise multiplication.

All modules output frame-wise multi-label posteriors for $C$ classes via a shared linear readout: $\hat{\mathbf{y}}_t = \sigma(\mathbf{W}_o \mathbf{h}_t + \mathbf{b}_o)$, where $\mathbf{W}_o \in \mathbb{R}^{C \times H}$ and $\mathbf{b}_o \in \mathbb{R}^{C}$ are the learnable readout parameters, and $\mathbf{h}_t$ denotes the hidden state vector produced by the temporal module.
As detailed in Sec.~\ref{sec:instantiations}, we categorize each method based on how it constrains these gating factors (fixed scalars vs. adaptive gates) and its parameter trainability.

\subsection{Frozen Pre-Trained CNN Backbone}
\label{sec:backbone}
We utilize the frame-wise MobileNet (fmn10) from EfficientSED \cite{morocutti2025exploring} as our frozen front-end.
This backbone is derived from MobileNetV3 by removing global pooling and adjusting strides to preserve temporal resolution.
Given an input audio clip, it produces a sequence of embeddings $\{\mathbf{z}_t\}_{t=1}^{T}$ with a resolution of 40 ms. For a 10-s clip, $T=250$.
We initialize the backbone with pre-trained weights and keep them frozen.

\subsection{Model Instantiations}
\label{sec:instantiations}
We evaluate four instantiations of the unified framework, ranging from memory-less baselines to complex gated architectures.
For the RNN variants (GRU, LSTM, ESN), we employ SED models under both unidirectional and bidirectional settings.
The bidirectional formulation follows the standard approach by running the dynamics (Eq.~(\ref{eq:unified_update})) in both forward and backward time directions and concatenating the resulting states $\mathbf{h}_t = [\mathbf{h}_t^{\text{fwd}}; \mathbf{h}_t^{\text{bwd}}]$ \cite{schuster1997bidirectional}.

\subsubsection{Linear Baseline}
The linear model serves as a non-temporal baseline, representing the case where no history is retained.
Within the unified framework, this can be viewed as a degenerate case where
\begin{itemize}
    \item $\mathbf{\Gamma}_r = \mathbf{0}$ (No retention), 
    \item $\mathbf{\Gamma}_u = \mathbf{1}, \quad \mathbf{\Gamma}_o = \mathbf{1}$. 
\end{itemize}
The internal state is effectively the instantaneous input projection $\mathbf{s}_t = \tilde{\mathbf{s}}(\mathbf{z}_t)$. The shared classifier is applied independently to each frame.

\subsubsection{Gated Recurrent Units}
\label{sec:gru}
The GRU \cite{cho2014learning} introduces adaptive gating, in which the retention and update factors are coupled and learned from data. In the unified framework, this corresponds to the set as follows:
\begin{itemize}
    \item $\mathbf{\Gamma}_u = \mathbf{u}_t$ (Update gate, computed by a neural network), 
    \item $\mathbf{\Gamma}_r = 1 - \mathbf{u}_t$ (Coupled retention), 
    \item $\mathbf{\Gamma}_o = \mathbf{1}$. 
\end{itemize}
This coupling forces the model to balance erasing past knowledge with writing new information.
We employ a stacked GRU with two layers and hidden size $H=256$ per direction, a configuration often found in standard CRNN baselines \cite{chen2024semi}.
Dropout with $p=0.3$ is applied between layers and before the final classification head to prevent overfitting.
We select the GRU as our primary benchmark because it represents the current state-of-the-art standard for temporal aggregation in SED. 

\subsubsection{Long Short-Term Memory}
\label{sec:lstm}
The LSTM \cite{hochreiter1997long} offers the most flexible dynamics within the framework, with fully independent gates and a separation between the memory cell $\mathbf{s}_t$ and the hidden output $\mathbf{h}_t$. In the unified framework, this is equivalent to the following setting:
\begin{itemize}
    \item $\mathbf{\Gamma}_r = \mathbf{f}_t$ (Forget gate, independent), 
    \item $\mathbf{\Gamma}_u = \mathbf{i}_t$ (Input gate, independent), 
    \item $\mathbf{\Gamma}_o = \mathbf{o}_t$ (Output gate). 
\end{itemize}
This decoupling allows the LSTM to retain memory over long durations without necessarily exposing it to the output at every step.
Similar to the GRU, we use a two-layer architecture with $H=256$ units and $p=0.3$ dropout.
We include the LSTM as a high-complexity baseline to determine whether its explicit memory management and larger parameter space yield a significant performance gain over the more streamlined GRU and ESN architectures, or if the added computational cost is redundant for this task.

\subsubsection{Echo State Network}
\label{sec:esn}
To push the boundaries of parameter efficiency beyond merely freezing the CNN feature extractor, we employ the ESN in the reservoir computing framework as a lightweight alternative to fully trainable RNN variants.
The ESN is a special case of an RNN in which the recurrent dynamics are fixed rather than learned \cite{jaeger2001echo}. A more general model is the leaky-integrator ESN, which enables flexible tuning of the reservoir timescale \cite{jaeger2007optimization}.
In the unified framework, this is characterized by constant scalar gating factors as follows: 
\begin{itemize}
    \item $\mathbf{\Gamma}_r = 1 - \alpha$, 
    \item $\mathbf{\Gamma}_u = \alpha, \quad \mathbf{\Gamma}_o = \mathbf{1}$. 
\end{itemize}
Here, $\alpha \in (0, 1]$ is a hyperparameter denoted as the leaking rate.
Crucially, unlike GRU/LSTM, the recurrent weights defining $\tilde{\mathbf{s}}$ are randomly initialized and frozen.
Only the final readout weights $\mathbf{W}_o$ are trained.
This setup tests the efficacy of fixed, random projections for temporal context.

To maximize the potential of the reservoir computing approach, we optimize the hyperparameters of the ESN and the readout training configuration using the Optuna framework \cite{akiba2019optuna}. We employ the Tree-structured Parzen Estimator (TPE) sampler to efficiently explore the configuration space over 150 trials for each training set size.
In the multi-class trigger SED setting, we conduct a single study using the full training set (6000 clips).
In the few-shot user-adaptive setting, we conduct a separate study for each support set size $K \in \{1,2,3,4,5\}$, where $K$ denotes the number of support clips.
The optimization objective is to maximize the PSDS1 score on the validation set.
The search space for the ESN dynamics includes the spectral radius $\rho$ (sampled uniformly from $0.1$ to $1.8$), the leaking rate $\alpha$ (uniformly from $0.05$ to $1.0$), and the input scaling factor $\sigma_{\text{in}}$ (log-uniformly from $0.01$ to $5.0$). For the supervised readout layer, we tuned the learning rate $\eta$ within a log-uniform range of $10^{-4}$ to $3 \times 10^{-3}$. Other structural parameters, such as the reservoir density ($d=0.1$) and the random sparse topology, are fixed to maintain a consistent reservoir complexity. For the bidirectional ESN variants, hyperparameters are shared across both the forward and backward reservoirs to streamline the search process.

We set the number of reservoir nodes to 1,024 in the multi-class trigger SED task to provide sufficient capacity for discriminating multiple overlapping categories (see Sec.~\ref{sec:experiments}).
For the few-shot user-adaptive trigger SED, we reduce the reservoir size to match the capacity of GRU/LSTM, mitigating overfitting under limited support data (Sec.~\ref{sec:experiments}).

\subsection{Evaluation Metrics}
\label{sec:metrics}
We evaluate model performance on misophonia trigger SED tasks (see Sec. \ref{sec:tasksetup}) using the macro-averaged F1 score (both segment-based and event-based) and the Polyphonic Sound Detection Score (PSDS) \cite{bilen2020framework}.
For the segment-based F1 score ($F1_{\text{seg}}$), predictions are evaluated on per-second segments.
The event-based F1 score ($F1_{\text{event}}$) serves as our strict measure for discriminative capability and fine-grained localization, counting a true positive only if the class is correct and the onset/offset matches within a 200 ms collar.
Following the DCASE Task 4 evaluation protocol, PSDS is commonly reported under two application scenarios, denoted as PSDS1 and PSDS2 \cite{ronchini2022benchmark}.
In this work, to assess detection robustness independent of a fixed decision threshold, we report PSDS1 (Scenario 1).
We select Scenario 1 ($\rho_{\text{DTC}}=0.7$, $\rho_{\text{GTC}}=0.7$, and $\alpha_{\text{CT}}=0$) over Scenario 2 because our application requires high temporal precision for short, transient triggers.
The strict overlap threshold of 0.7 ensures that detected events are well-localized in time, while cross-trigger confusion is effectively monitored by the strict classification constraints of $F1_{\text{event}}$.
All metrics are normalized to $[0, 1]$.

\subsection{Optimization and Training Details}
\label{subsec:optimization}
As emphasized in Sec.~\ref{sec:backbone}, the CNN backbone remains frozen; gradient updates are applied only to the temporal modules described in Sec. \ref{sec:instantiations}.
For trainable modules (Linear/GRU/LSTM), we optimized all module parameters end-to-end using Adam \cite{kingma2014adam} with a learning rate of $\eta = 1\times 10^{-4}$ and a batch size of 1024.
For the ESN modules, only the linear readout parameters are updated; the readout learning rate is selected by Optuna (Sec.~\ref{sec:esn}), and the batch size is kept identical for fair comparison.

We use the validation set to select all inference-time hyperparameters.
First, we select the training checkpoint with the best validation PSDS1 (Scenario 1).
Second, we tune the median-filter window size on the validation set to maximize PSDS1, to suppress short spurious activations and reduce over-segmentation.
Finally, for F1 reporting, we determine class-wise decision thresholds on the validation set by maximizing the event-based F1, $F1_{\text{event}}$, and compute $F1_{\text{seg}}$ using the same thresholds.
All models are implemented in PyTorch~\cite{paszke2019pytorch}.

\section{Experiments}
\label{sec:experiments}
We evaluate our approach under two complementary settings:
(i) Multi-class trigger SED on synthetic soundscapes, and
(ii) Few-shot user-adaptive trigger SED that models a personalization scenario for misophonia.
Hereafter, we refer to the CNN-based hybrid model by the name of its subsequent temporal module. 

\subsection{Tasks and Setup}
\label{sec:tasksetup}
\subsubsection{Multi-Class Trigger SED}
In this task, we evaluate the performance of the hybrid models on the seven-class trigger SED problem.
We utilize the synthetic dataset constructed in Sec.~\ref{sec:dataset}, employing the fixed train/validation/test splits (6000/2000/2000 clips) to ensure reproducibility.
The primary goal is to assess how well different temporal modules discriminate and localize the oronasal and mechanical trigger sounds under the unified framework described in Sec.~\ref{sec:methods}.
Specifically, we compare both unidirectional and bidirectional variants of ESNs, GRUs, and LSTMs against the linear baseline.
In addition to detection accuracy, we analyze the number of trainable parameters for each model to evaluate the trade-off between performance and model complexity.
Evaluation is performed using the metrics defined in Sec.~\ref{sec:metrics}, specifically focusing on PSDS1 to measure time-localized detection performance, alongside $F1_{\text{event}}$ and $F1_{\text{seg}}$.

\subsubsection{Few-Shot User-Adaptive Trigger SED}
To simulate a personalization scenario where a user configures the system for a specific trigger, we conduct a few-shot detection experiment focusing on the ``eating sound'' class.
We select clips containing distinct chewing or smacking sounds to represent a specific user's trigger profile.
Adopting a support-query protocol inspired by DCASE Task 5 \cite{morfi2021few}, we vary the number of support clips (shots), with $K \in \{1, 2, 3, 4, 5\}$.
We specifically compare BiGRU and BiESN as representative models for few-shot personalization.
For each $K$, we adapt the BiGRU and BiESN models (defined in Sec.~\ref{sec:methods}) using only the $K$ provided support clips, utilizing non-target segments within these clips as negative examples.
While we utilize the same model architectures as in the multi-class task, we set the reservoir size of the BiESN to $N=256$, as preliminary experiments indicated that this capacity is sufficient for the binary detection task.
We define a fixed support pool of eight curated eating-sound recordings. For each run, we sample $K$ support clips from this pool using a different random seed, resulting in different support combinations across runs. We report the mean and standard deviation over 10 runs (different seeds) for each $K$.

\subsection{Results}
\subsubsection{Multi-Class Trigger SED}
\label{sec:multiclass_results}
Table~\ref{tab:main_results} summarizes detection performance and trainable parameter counts for the seven-class trigger SED task. Bidirectional temporal modeling consistently improves performance across temporal modules: PSDS1 increases from 0.55 to 0.63 for GRU, from 0.52 to 0.60 for LSTM, and from 0.49 to 0.55 for ESN, indicating that leveraging both past and future context within each 10-s clip is beneficial. Among all methods, BiGRU achieves the best overall performance ($\text{PSDS1} = 0.63,\, F1_{\text{event}} = 0.65,\, F1_{\text{seg}} = 0.85$), while BiLSTM follows closely ($\text{PSDS1} = 0.60,\, F1_{\text{event}} = 0.64,\, F1_{\text{seg}} = 0.84$).

Despite its substantially smaller trainable footprint, BiESN provides competitive accuracy. Specifically, BiESN attains $\text{PSDS1} = 0.55$ ($F1_{\text{event}} = 0.57$), achieving performance comparable to the unidirectional GRU ($\text{PSDS1} = 0.55,\, F1_{\text{event}} = 0.56$) while requiring only 14,343 trainable parameters, which is approximately 72$\times$ fewer than the GRU (1,037,319 parameters). Overall, BiESN offers a favorable trade-off between detection capability and model complexity when training cost and model size are constrained, at the expense of a moderate reduction in peak performance compared with fully trainable temporal modules.

\begin{table}[t]
\centering
\caption{Performance on Multi-Class Trigger SED (seven classes).}
\label{tab:main_results}
\renewcommand{\arraystretch}{1.1}
\setlength{\tabcolsep}{4pt}
\begin{tabular}{l c c c c r}
\hline
\multicolumn{1}{c}{Model} &
\multicolumn{1}{c}{Dir.} &
\multicolumn{1}{c}{PSDS1} &
\multicolumn{1}{c}{\makecell{Event\\F1}} &
\multicolumn{1}{c}{\makecell{Segment\\F1}} &
\multicolumn{1}{c}{\makecell{Trainable\\params}} \\
\hline
Linear & -- & 0.47 & 0.52 & 0.80 & 6,727 \\

GRU & \makecell{Uni\\Bi} &
\makecell{0.55\\0.63} &
\makecell{0.56\\0.65} &
\makecell{0.81\\0.85} &
\makecell[r]{1,037,319\\2,221,831} \\

LSTM & \makecell{Uni\\Bi} &
\makecell{0.52\\0.60} &
\makecell{0.55\\0.64} &
\makecell{0.81\\0.84} &
\makecell[r]{1,300,487\\2,879,239} \\

ESN & \makecell{Uni\\Bi} &
\makecell{0.49\\0.55} &
\makecell{0.52\\0.57} &
\makecell{0.78\\0.81} &
\makecell[r]{7,175\\14,343} \\
\hline
\end{tabular}
\end{table}

\subsubsection{Few-Shot User-Adaptive Trigger SED}
\begin{figure}[!t]
  \centering
  \includegraphics[width=\linewidth]{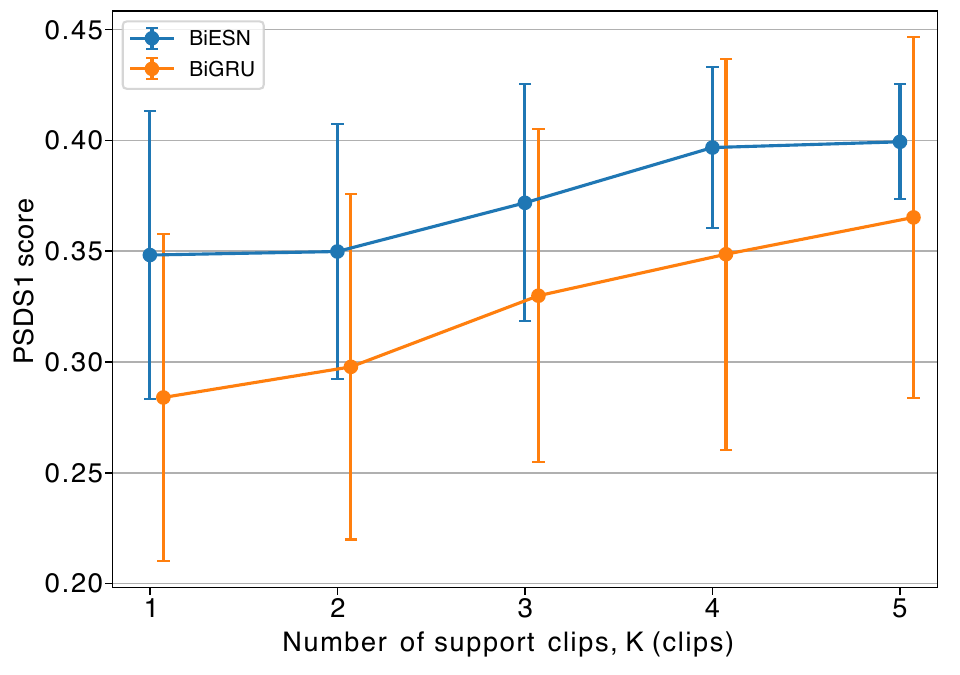}
  \vspace{1mm}
  \caption{Few-shot user-adaptive trigger SED performance measured by PSDS1 as a function of the number of support clips $K$. In each run, $K$ support clips are sampled from a fixed pool of eight recordings using a different random seed. Points indicate the mean over 10 runs, and error bars denote $\pm 1$ standard deviation.}
  \label{fig:fewshot}
\end{figure}

Fig.~\ref{fig:fewshot} plots PSDS1 as a function of the number of support clips $K$ for the user-adaptive ``eating sound'' detection task. BiESN achieves consistently higher mean PSDS1 than BiGRU across all $K$ in this setting, while BiGRU exhibits larger variability across runs, suggesting a stronger sensitivity to the particular support selection and training stochasticity under limited data. Overall, the results indicate that readout-only adaptation with a fixed reservoir provides more reliable few-shot performance under our constrained personalization setup.

\section{Conclusion}
In this study, we investigated misophonia trigger SED as a fundamental step toward on-device applications designed to mitigate the adverse effects of misophonia.
To address the lack of standardized datasets in this domain, we curated a strongly labeled dataset via soundscape synthesis, which is one of our main contributions. This approach enabled the generation of precise onset/offset annotations at scale, overcoming the difficulty of collecting large real-world datasets for subtle mouth/nasal triggers.
Using this dataset, we evaluated CNN-based hybrid SED models, conducting experiments in both multi-class and few-shot user-adaptive settings to explore practical personalization.
A critical trade-off was identified regarding model architecture: while the BiGRU achieved the highest detection accuracy, the BiESN offered competitive performance with orders-of-magnitude fewer trainable parameters and simpler readout-only training.
This finding is particularly significant for future personalized, real-time misophonia mitigation techniques. 

Despite these promising results, challenges remain in bridging the gap between synthetic training and real-world deployment.
Since the current performance relies on synthetic soundscapes, robustness against the diverse acoustic conditions of real environments must be verified.
Future work will address the synthetic-to-real domain gap through domain adaptation and evaluate system latency for streaming inference.
Moreover, given the highly individualized nature of misophonia, we aim to extend our framework toward user-adaptive detection (few-shot personalization) and fine-grained intra-class modeling.
For instance, enabling the system to discriminate specific acoustic patterns such as ``slurping'' and ``crunching'' within the broader ``eating'' class is crucial, as a user may be triggered by one but not the other.
This approach moves us closer to providing practical, everyday relief tailored to individual needs.
\bibliographystyle{IEEEtran}
\bibliography{references}
\end{document}